\documentclass[12pt]{article}
\begin{document}
\pagestyle{plain}
\title{Electron spin motion in the delta-function  pulse}
\author{Miroslav Pardy\\
Department of Physical Electronics \\
Masaryk University \\
Kotl\'{a}\v{r}sk\'{a} 2, 611 37 Brno, Czech Republic\\
e-mail:pamir@physics.muni.cz}
\date{\today}
\maketitle
\vspace{30mm}

\begin{abstract}
We formulate the Bargman-Michel-Telegdi  (BMT) equation for electron spin motion in a plane wave and in the Dirac delta-function pulse.
We compare the BMT solution with the Wolkow solution of the Dirac equation.
 The Wolkow solution for the spin is not 
identical with the solution following from the BMT .

\end{abstract}

\vspace{30mm}

\baselineskip 15 pt

\section{Introduction}

The problem of interaction an elementary particles with the laser field is,
at present time, one of the most prestige problem in the particle
physics. It is supposed that, in the future, the laser will
play the same role in particle physics as the linear or circular accelerators working in today particle laboratories.
The lasers nowadays provide one of the most powerful
sources of electromagnetic (EM) radiation under
laboratory conditions and thus inspire the fast growing
area of high field science aimed at the exploration of
novel physical processes (Mourou et al. 2006; Marklund et al., 2006; Salamin et al., 2006). 

Lasers have already demonstrated
the capability to generate light with the intensity
of ${\rm 2\times 10^{22}W/cm^{2}}$ (Yanovsky, et al., 2008) and projects to achieve ${\rm 10^{26} W/cm^{2}}$ (Dunne, 2006)
are under way. Further intensity growth towards and
above ${\rm 10^{23}W/cm^{2}}$ will bring us to experimentally unexplored
regimes. 

Here, we consider the interaction of an
electron with a plane wave field or, with a ultrashort (Dirac $\delta$-function) laser pulse. The quantum motion of electron in 
a plane wave was firstly described by Wolkow (1935). It involves the classical limit with the classical solution.

First, we consider the classical approach to motion of a charged particle in a plane  field and then in a Dirac $\delta$-function  pulse which is the final goal of  attosecond  laser pulses   (Agostini et al.,2004).

Then we formulate  the Bargman-Michel-Telegdi equation for electron spin motion in a plane wave and in the laser pulse.
We compare the solution with the solution which was derived from the Wolkow solution of the Dirac equation for zero and the anomalous magnetic moment of particle with spin one half.

\section{Classical interaction of a charged particle with a plane wave}

To find motion of an electron in a periodic electromagnetic field, it is suitable to solve Lorentz equation
in general with four potential $A_{\mu} = a_{\mu}A(\varphi)$, where $\varphi = kx, k^{2} = 0$.
Following Meyer (1971) we apply his method and then we are prepared to consider radiation
reaction which has some influence on the motion of electron in the
electromagnetic field.

The Lorentz equation with $A_{\mu} = a_{\mu}A(\varphi)$ reads:

$$\frac {dp_{\mu}}{d\tau} = \frac {e}{m}F_{\mu\nu}p^{\nu}
= \frac {e}{m}(k_{\mu}a\cdot p - a_{\mu}k\cdot p)A'(\varphi); \quad A' = \frac{dA}{d\varphi}, 
\eqno(1)$$
where $\tau$ is proper time and $p_{\mu} = m(dx_{\mu}/d\tau)$.
After multiplication of the last equation by $k^{\mu}$, we get with regard to
the Lorentz condition $ 0 = \partial_{\mu}A^{\mu} =
a^{\mu}\partial_{\mu}A(\varphi) = k_{\mu}a^{\mu}A'$, or,
$k\cdot a = 0$ and $k^{2} = 0$, the following equation:

$$\frac {d(k\cdot p)}{d\tau} = 0\eqno(2)$$
and it means that $k\cdot p$ is a constant of the motion and it can be defined
by the initial conditions  for instance at time $\tau = 0$. If we put
$p_{\mu}(\tau = 0) = p_{\mu}^{0}$, then we can
write $k\cdot p = k\cdot p^{0}$. At this moment we have with $d\varphi = k\cdot dx$:

$$k\cdot p = \frac{mk\cdot dx}{d\tau} = m\frac {d \varphi}{d\tau},
\eqno(3)$$
or,

$$\frac{d\varphi}{d\tau} = \frac {k\cdot p^{0}}{m}.
\eqno(4)$$

So, using the last equation and relation $d/d\tau =
(d/d\varphi)d\varphi/d\tau$, we can write equation (1) in the form
$(dp_{\mu}/d\tau  =(d\varphi/d\tau)(d p_{\mu}/d\varphi))$:

$$\frac {dp_{\mu}}{d\varphi} = \frac {e}{k\cdot p^{0}}
(k_{\mu}a\cdot p - a_{\mu}k\cdot p^{0})A'(\varphi) = 
e\left(k_{\mu}\frac{a\cdotp}{k\cdotp^{0}} - a_{\mu}\right)A'(\varphi)\eqno(5)$$
giving (after multiplication by $a^{\mu}$)

$$\frac {d(a\cdot p)}{d\varphi} = - ea^{2}A' ,\eqno(6)$$
or,

$$a\cdot p = a\cdot p^{0} - ea^{2}A.\eqno(7)$$

Substituting the last formula into (5), we get:

$$\frac {dp_{\mu}}{d\varphi} = -e\left(a_{\mu} -
\frac {k_{\mu}a\cdot p^{0}}{k\cdot
p^{0}}\right)\frac {dA}{d\varphi} -
\frac {e^{2}a^{2}}{2k\cdot p^{0}}\frac {d(A^{2})}{d\varphi}
k_{\mu}.
\eqno(8)$$

This equation  can be immediately integrated to
give the resulting momentum in the
form:

$$p_{\mu} = p_{\mu}^{0} - e\left(
A_{\mu} - \frac {A^{\nu}p_{\nu}^{0}k_{\mu}}{k\cdot p^{0}}\right)
- \frac {e^{2}A^{\nu}A_{\nu}k_{\mu}}{2k\cdot p^{0}}.\eqno(9)$$

\section{Classical interaction of a charged particle with a $\delta$-function pulse}

One of the primary goals of ultrashort laser science is to provide more insights into the dynamics of atomic electrons. One general interest is the direct probing in time of hyperfast electronic rearrangements following the creation of an inner-shell
hole. There is a  study using sub-femtosecond burst of XUV light
probed the motion of an electron wave packet under the influence of
an infrared laser's electric field. Furthermore, the precise timing of the electron wave packet emitting the high harmonics can be measured by observing the two-photon ionization electron
energy spectrum These pioneering experiments are
reviewed  by Agostini et al. (2004). 

What are the limits and future of attosecond pulses? 
The goal of the laser physics is to generate very short pulses. At
present time we are able to generate the  attosecond pulses (Agostini et al., 2004). Nevertheless, the final goal of short pulse laser laboratories is to generate laser pulse in the  $\delta$-function form, and there is no theory which restricts the attainability of such pulses. It is not excluded that the mystery of the Higgs boson will be revealed just using such laser pulses. So, let us first remember the rigorous theory of $\delta$-function.

The $\delta$-function mathematical theory can be presented in a very simple way (Martynenko, 1973).  Namely, using the definition of the unite Heaviside step function denoted as the $\eta$-function. It is defined by
the relation:

$$\eta(t) = \left\{ \begin{array}{c} 0, \quad t <  0  \\
1, \quad t \geq 0
\end{array} \right. .
\eqno(10)$$

The $\eta$-function is the limiting case of the sequence  $\eta_{n}(t)$.

$$\eta_{n}(t) = \frac{1}{2} + \frac{1}{\pi}\arctan(nt); \quad
|\arctan(nt)| < \frac{\pi}{2}. \eqno(11)$$

Using this sequence, we define the $\delta$-generating sequence 

$$\delta_{n}(t) = \frac{d}{dt}\eta_{n}(t),\eqno(12)$$ 
or, 

$$\delta_{n}(t) = \frac{n}{\pi(t^{2}n^{2} + 1)}. \eqno(13)$$ 

The $\delta$-function is then defined as the limiting case of the last relation 

$$\delta(t) = \lim_{n\rightarrow \infty} \frac{n}{\pi(t^{2}n^{2} + 1)}.\eqno(14)$$ 

So, we get that the $\delta$-function is derived as

$$\delta(t) = \left\{ \begin{array}{c} 0, \quad t \neq  0  \\
\infty , \quad t = 0
\end{array} \right..
\eqno(15)$$ 

If we perform the integration of the $\delta_{n}$ function, we get:

$$\int_{-\infty}^{\infty}\delta_{n}(t)dt = \int_{-\infty}^{\infty}\frac{n}{\pi(t^{2}n^{2} + 1)}dt = \frac{2}{\pi}\int_{0}^{\infty}d(\arctan(nt)) = 1 .\eqno(16)$$

For $t = 0$, we have

$$\delta_{n}(t) = \frac{n}{\pi}; \quad \delta(t) = 
\lim_{n\rightarrow \infty}\delta_{n}(t)\eqno(17)$$
and 

$$ \int_{-\infty}^{\infty}\delta(t)dt =  1.  \eqno(18)$$

So, we write

$$\frac{d}{dt}\eta(t) = \delta(t) = \left\{ \begin{array}{c} 0, \quad t \neq  0  \\
\infty , \quad t = 0
\end{array} \right.;\quad \int_{-\infty}^{\infty}\delta(t)dt = 1.
\eqno(19)$$ 

Let us remark that the $\delta$-function was in the history of
mathematics used also by Poisson, Cauchy, Hermite and others. At
present time the $\delta$-function is called the Dirac $\delta$-function
because it was introduced into quantum mechanics rigorously by Dirac.

The $\delta$-function has also meaning in classical mechanics.
Newton's second law
for the interaction of a massive particle with mass $m$ with an impact force $\delta(t)$
is as follows:

$$m\frac{d^{2}x}{dt^{2}} =  P\delta(t),\eqno (20)$$
where P is some constant. If we express $\delta$-function by the relation $\delta(t) =\dot\eta(t)$,
then from eq. (20) $\dot x(t)= P/m$ follows immediately. The physical meaning of the quantity $P$ can be deduced from equation $F = P\delta(t)$. After $t$-integration we have $\int F dt  = \int m(dv/dt)dt = mv =  P$, where $m$ is mass of a body and $v$ its final velocity (with $v(0)= 0$). It means that the value of $P$ can be determined a posteriori and then this value can be used in more complex equations than eq. (20). Of course it is
necessary to suppose that $\delta$-form of the impact force is adequate approximation of the experimental situation.

Now, if we put into formula (9) the four-potential $A_{\mu} = a_{\mu}A(\varphi) = a_{\mu}\eta(\varphi)$
of the impact force,
then for $\varphi\ > 0$ when $\eta > 1$, we get:

$$p_{\mu} = p_{\mu}^{0} - e\left(
a_{\mu} - \frac {a^{\nu}p_{\nu}^{0}k_{\mu}}{k\cdot p^{0}}\right)
- \frac {e^{2}a^{\nu}a_{\nu}k_{\mu}}{2k\cdot p^{0}}.
\eqno(21)$$

The last equation can be used to determination of the magnitude of $a_{\mu}$
similarly as it was done in discussion to the eq. (20). It can be evidently expressed as the number of $k$-photons in electromagnetic momentum. For
$\varphi < 0$, it is $\eta = 0$ and therefore $p_{\mu} = p_{\mu}^{0}$

It is still necessary to say what is the practical realization of the
$\delta$-form potential. We know from the Fourier analysis
that the Dirac $\delta$-function can be expressed by integral
in the following form:

$$\delta(\varphi) = \frac {1}{\pi}\int_{0}^{\infty}\cos(s\varphi)ds.
\eqno(22)$$

So, the $\delta$-force and $\delta$-potential can be realized as the continual
superposition of the harmonic waves. In case it will be not possible
to realize  experimentally it,
we can approximate the integral formula by the summation
formula as follows:

$$\delta(\varphi) \approx \frac {1}{\pi}\sum_{0}^{\infty}\cos(s\varphi).
\eqno(23)$$

If we consider the $\delta$-form electromagnetic pulse, then we can write

$$F_{\mu\nu} = a_{\mu\nu}\delta(\varphi).\eqno(24)$$
where $\varphi = kx = \omega t - {\bf k}{\bf x}$. In order to obtain the
electromagnetic impulsive force in this form, it is necessary
to define the four-potential in the following form:

$$A_{\mu} = a_{\mu}\eta(\varphi),\eqno(25)$$
where function $\eta$ is the Heaviside unit step  function defined by
the relation:

$$\eta(\varphi) = \left\{ \begin{array}{c} 0, \quad \varphi <  0  \\
1, \quad \varphi \geq 0
\end{array} \right. .
\eqno(26)$$

If we define the four-potential by the equation (25), then the
electromagnetic tensor with impulsive force is of the form:

$$F_{\mu\nu} = \partial_{\mu}A_{\nu} - \partial_{\nu}A_{\mu} =
(k_{\mu}a_{\nu} - k_{\nu}a_{\mu})\delta(\varphi) =
a_{\mu\nu}\delta(\varphi).\eqno(27)$$

\section{Motion of the spin-vector in electromagnetic field}

Bargmann, Michel and Telegdi  (BMT) (Berestetzkii et al., 1989) derived so called BMT equation for motion of spin in the electromagnetic field, in the form 

$$\frac{da_{\mu}}{ds} = \alpha F_{\mu\nu}a^{\nu} + \beta v_{\mu}F^{\nu\lambda}v_{\nu}a_{\lambda},\eqno(28)$$
where $a_{\mu}$ is so called axial vector describing the classical
spin, $v_{\mu}$  is velocity and  constants $\alpha$ and $\beta$ were determined after the comparison of the postulated equations with the 
non-relativistic quantum mechanical limit. The result of such comparison is the final form of so called BMT equations:

$$\frac{da_{\mu}}{ds} = 2\mu F_{\mu\nu}a^{\nu} -2\mu'v_{\mu}F^{\nu\lambda}v_{\nu}a_{\lambda},\eqno(29)$$
where $\mu$ is magnetic moment of electron following directly from the
Dirac equation and $\mu'$ is anomalous magnetic moment of electron
which can be calculated as the radiative correction to the interaction
of electron with electromagnetic field and it 
follows from quantum electrodynamics.

 The BMT equation has more earlier origin. The first attempt to describe the spin motion in electromagnetic field using the special theory of relativity was performed by Thomas (1926). However, the basic ideas on the spin motion was established by Frenkel (1926, 1958). After appearing the Frenkel basic article,
many authors published the articles concerning the spin motion (Ternov
et al., 1980; Tomonaga, 1997; Ohanian, 1986). At present time, 
spin of electron is considered as its  physical attribute 
which follows only from the Dirac equation.

It was shown by Rafanelli and Schiller (1964), (Pardy, 1973) 
that the BMT equation can be derived from the classical limit, i.e. from the WKB solution of the Dirac equation with the anomalous magnetic moment.

If we introduce the average value of the vector of spin in the rest system by the quantity $\mbox {\boldmath $\zeta$}$,
 then the 4-pseudovector  $a^{\mu}$  is of the from 
$a^{\mu} = (0, \mbox {\boldmath $\zeta$})$ (Berestetzkii et al., 1989;
 Pardy, 2012). The momentum four-vector of a particle is $p ^{\mu} = (m, 0)$ in the rest system of a particle. Then the equation $a^{\mu}p_{\mu} = 0 $
is valid not only in the rest system of a particle but in the arbitrary system as a consequence of the relativistic invariance. The following general formula is also valid in the arbitrary  system
$a^{\mu}a_{\mu} = - \mbox {\boldmath $\zeta$}^{2}$.

The components of the axial 4-vector $a^{\mu}$ in the reference system, where particle is moving with the velocity ${\bf v} = {\bf p}/\varepsilon$ can be obtained by application of the Lorentz transformation to the rest system and they are as follows (Berestetzkii et al., 1989):

$$a^{0} = \frac{|{\bf p}|}{m}\mbox {\boldmath $\zeta$}_{\parallel}, \quad {\bf a}_{\perp} = \mbox {\boldmath $\zeta$}_{\perp}, \quad a_{\parallel} = \frac{\varepsilon}{m}\mbox {\boldmath $\zeta$}_{\parallel}, \eqno(30)$$
where suffices $\parallel, \perp$ denote the components of ${\bf a}$, $\mbox {\boldmath $\zeta$}$ parallel and perpendicular to the direction ${\bf p}$. The formulas for the spin components can be also rewritten in the more compact form as follows (Berestetzkii et al., 1989):

$${\bf a} = {\mbox {\boldmath $\zeta$}} + \frac{{\bf p}({\mbox {\boldmath $\zeta$}}{\bf p})}{m(\varepsilon + m)}, \quad a^{0} = \frac{{\bf a}{\bf p}}{\varepsilon} = \frac{{\mbox {\boldmath $\zeta$}}{\bf p}}{m}, \quad
{\bf a}^{2} = {\mbox {\boldmath $\zeta$}}^{2} + \frac{({\bf p}{\mbox {\boldmath $\zeta$}})^{2}}{m^{2}}.\eqno(31)$$

The equation for the change of polarization can be obtained immediately from the BMT equation in the following form (Berestetzkii et al., 1989):

$$ \frac{d{\bf a}}{dt} = \frac{2\mu m}{\varepsilon}{\bf a}\times{\bf H} + \frac{2\mu m}{\varepsilon}({\bf a}{\bf v}){\bf E} - \frac{2\mu' \varepsilon}{m}{\bf v}({\bf a}{\bf E})\; + $$

$$+ \frac{2\mu'\varepsilon}{m}{\bf v}({\bf v}({\bf a}\times {\bf H})) +  \frac{2\mu'\varepsilon}{m}{\bf v}({\bf a}{\bf v})({\bf v}{\bf E}), \eqno(32)$$
where we used the relativistic relations $c =1$, $ds = dt\sqrt{1 - v^{2}}$ , $\varepsilon = m\sqrt{1 - v^{2}}$ and the following components of the electromagnetic field (Landau et al., 1988):

$$F^{\mu\nu} = \left(\begin{array}{cccc}
0 & -E_{x} & -E_{y} & -E_{z}\\
E_{x} & 0 & -H_{z} & H_{y}\\
E_{y} & H_{z} & 0 & -H_{x}\\
E_{z} & -H_{y} & H_{x} & 0\\
\end{array} \right) \stackrel {d}{=} ({\bf E}, {\bf H}); \quad F_{\mu\nu} = ({-\bf E }, {\bf H}).\eqno(33)$$

Inserting equation ${\bf a}$ from eq. (31) into  eq. (32) and using equations 

$${\bf p} = \varepsilon{\bf v}, \quad \varepsilon^{2} = {\bf p}^{2} + m^{2}, \quad \frac{d{\bf p}}{dt} = e{\bf E} + e({\bf v}\times {\bf H}),\quad
\frac{d\varepsilon}{dt} = e({\bf v}{\bf E}),\eqno(34)$$
we get after long but simple mathematical operations the following equation for the polarization $\mbox {\boldmath $\zeta$}$

$$\frac{d {\mbox {\boldmath $\zeta$}}}{dt} = 
\frac{2\mu m  + 2\mu'(\varepsilon - m)}{\varepsilon}{\mbox {\boldmath $\zeta$}}\times {\bf H} \quad + $$

$$\frac{2\mu' \varepsilon}{\varepsilon + m}({\bf v}{\bf H})({{\bf v}\times \mbox {\boldmath $\zeta$}}) + \frac{2\mu m  + 2\mu' \varepsilon}{\varepsilon + m}{\mbox {\boldmath $\zeta$}}\times ({\bf E}\times {\bf v}).\eqno(35)$$

The equation of motion of spin in electric field as far as first
order terms in velocity $v$ is obtained from eq. (32) in the form 

$$\frac{d {\mbox {\boldmath $\zeta$}}}{dt} =
(\mu  + \mu'){\mbox {\boldmath $\zeta$}}\times ({\bf E}\times {\bf  v}) = 
\left(\frac{e}{2m} + 2\mu'\right)
{\mbox {\boldmath $\zeta$}}\times ({\bf E}\times {\bf  v}).\eqno(36)$$

\section{BMT equation in the delta-function pulse}

Bargmann-Michel-Telegdi equation (BMT equation) is equation derived in 1959
For spin motion in electromagnetic field. If we denote the axial vector
describing spin as $S_{\mu}$, then the BMT equation reads:

$$\frac {dS_{\mu}}{d\tau} = 2\mu F_{\mu\nu}S^{\nu} - 2\left(\mu - \frac
{e\hbar}{2mc}\right)v_{\mu}F_{\nu\lambda}v^{\nu}S^{\lambda}\eqno(37)$$
where $\mu$ is the magnetic moment of spinning particle.

In case of the periodic magnetic field $A_{\mu} = a_{\mu}A(\varphi)$, which
gives $F_{\mu\nu} = (k_{\mu}a_{\nu} - k_{\nu}a_{\mu})A'(\varphi)$,
we have for BMT equation the following form:

$$\frac {dS_{\mu}}{d\tau} = 2\mu(k_{\mu}a_{\nu} - k_{\nu}a_{\mu})
A'(\varphi)S^{\nu} - 2\left(\mu - \frac{e\hbar}{2mc}\right)v_{\mu}
(k_{\nu}a_{\lambda} - k_{\lambda}a_{\nu})
A'(\varphi)v^{\nu}S^{\lambda}\eqno(38)$$

In case of the $\delta$-function pulse it is $A_{\mu} = a_{\mu}\eta'(\varphi) = a_{\mu}\delta(\varphi)$, or $A' = \delta(\varphi)$ and then we get fro eq. (35) using the formula $d/d\tau = d\varphi/d\tau(d/d\varphi) = kv(0)(d/d\varphi)$ 
 the following equation:

 $$\frac {dS_{\mu}}{d\varphi} = \frac {1}{kv(0)}
\left\{2\mu(k_{\mu}a_{\nu} - k_{\nu}a_{\mu})
\delta(\varphi)S^{\nu} - 2\left(\mu - \frac{e\hbar}{2mc}\right)v_{\mu}
(k_{\nu}a_{\lambda} - k_{\lambda}a_{\nu})
\delta(\varphi)v^{\nu}S^{\lambda}\right\}\eqno(39)$$

Now, let us approach the solution of the last equation. With the elementary knowledge of the properties of the Dirac $\delta$-function $\int f(x)\delta(x)dx = f(0)$ we get

$$S_{\mu}({\varphi}) = \frac {1}{kv(0)}
\left\{2\mu(k_{\mu}a_{\nu} - k_{\nu}a_{\mu})
S^{\nu}(0) - 2\left(\mu - \frac{e\hbar}{2mc}\right)v_{\mu}
(k_{\nu}a_{\lambda} - k_{\lambda}a_{\nu})
v^{\nu}S^{\lambda}(0)\right\}\eqno(40)$$
 
\section{Wolkow solution of the Dirac equation with Heaviside four-potential}

We know that the four-potential is inbuilt in the Dirac equation and we
also know that if the potential is dependent on $\varphi$, then,
there is explicit solution of the Dirac equation which  was found by
Wolkow (1935) and which is called Wolkow solution. The quantum mechanical
problem is to find solution of the Dirac equation with the $\delta$-form
four-potential (25) and from this solution determine the quantum motion of
the charged particle under this potential. Let us first remember the
Wolkow solution of the Dirac equation

$$(\gamma(p-eA) - m)\Psi = 0. \eqno(41)$$

Wolkow (1935) found the explicit solution of this equation for four-potential
$A_{\mu} = A_{\mu}(\varphi)$, where $\varphi = kx$. His solution
is of the form (Berestetzkii et al., 1989):

$$\Psi_{p} = R \frac {u}{\sqrt{2p_{0}}}e^{iS}  =
\left[1 + \frac {e}{2(kp)}(\gamma k)(\gamma A)\right]
\frac {u}{\sqrt{2p_{0}}}e^{iS},
\eqno(42)$$
where $u$ is an electron bi-spinor of the corresponding Dirac equation

$$(\gamma p - m)u = 0.
\eqno(43)$$

The mathematical object $S$ is the classical Hamilton-Jacobi function,
which  was determined in the form:

$$S = -px - \int_{0}^{kx}\frac {e}{(kp)}\left[(pA) - \frac {e}{2}
A^{2}\right]d\varphi.
 \eqno(44)$$

If we write Wolkow  wave function $\Psi_{p}$ in the form (42), then, for the
impulsive vector potential (25) we have:

$$S = -px - \left[e\frac {ap}{kp} - \frac {e^{2}}{2kp}a^{2}\right]\varphi,
\quad R = \left[1 + \frac {e}{2kp}(\gamma k)(\gamma a)\eta(\varphi)\right].
\eqno(45)$$

Our goal is to determine acceleration generated by the electromagnetic field
of the $\delta$-form which means that the four-potential $A_{\mu}$
is the Heaviside step function (10).
To achieve this goal, let us define current density
(Berestetzkii et al., 1989) as follows:

$$j^{\mu} = {\bar \Psi}_{p}\gamma^{\mu}\Psi_{p},
\eqno(46)$$
where $\bar\Psi$ is defined as the transposition of (42), or,

$$\bar\Psi_{p} = \frac {\bar u}{\sqrt{2p_{0}}}\left[1 +
\frac {e}{2(kp)}(\gamma A)(\gamma k)\right]
e^{-iS}.
\eqno(47)$$

After insertion of $\Psi_{p}$ and $\bar\Psi_{p}$
into the current density, we have with $A_{\mu} = a_{\mu}\eta(\varphi),
\eta^{2} = \eta$:

$$j^{\mu} = \frac {1}{p_{0}}\left\{p^{\mu} - ea^{\mu} +
k^{\mu}\left(\frac {e(pa)}{(kp)} - \frac {e^{2}a^{2}}{2(kp)}\right)
\right\}.
\eqno(48)$$
for $\eta > 0$, which is evidently related to eq. (21).

The so called kinetic momentum corresponding to $j^{\mu}$ is as follows
(Berestetzkii et al., 1989):

$$J^{\mu} = \Psi^{*}_{p}(p^{\mu} - eA^{\mu})\Psi_{p})
= {\bar \Psi}_{p}\gamma^{0}(p^{\mu} - eA^{\mu})\Psi_{p}) = $$

$$\left\{p^{\mu} - eA^{\mu} +
k^{\mu}\left(\frac {e(pA)}{(kp)} - \frac {e^{2}A^{2}}{2(kp)}\right)
\right\} + k^{\mu}\frac
{ie}{8(kp)p_{0}}F_{\alpha\beta}(u^{*}\sigma^{\alpha\beta}u),
\eqno(49)$$
where

$$\sigma^{\alpha\beta} = \frac {1}{2}(\gamma^{\alpha}\gamma^{\beta} -
\gamma^{\beta}\gamma^{\alpha}).\eqno(50)$$

Now, we express the four-potential by the step function.
In this case the kinetic momentum contains the tensor $F_{\mu\nu}$
involving $\delta$-function.
It means that there is a singularity at point $\varphi = 0$. This
singularity plays no role in the situation for
$\varphi > 0$ because in this case the $\delta$-function is zero. Then, the
kinetic momentum is the same as $j^{\mu}$.

\section{Spin motion of electron from the Wolkow solution}

In case of the Wolkow solution of the Dirac equation, the mathematical object which describes spin is as follows:

$$S_{\mu} = i\bar \psi\gamma_{5}\gamma_{\mu}\psi.\eqno(51)$$ 

After insertion of eqs. (42) and (47) into the last equation we get with $\widehat{k} = \gamma k, \widehat{A} = \gamma A,$ the following formula:
 
$$S_{\mu} = i\bar \psi\gamma_{5}\gamma_{\mu}\psi = i\frac{\bar u}{2p_{0}}
\left[\gamma_{5}\gamma_{\mu} + \frac {e}{2kp}\gamma_{5}\gamma_{\mu}
\widehat{A}\widehat{k} + \frac {e}{2kp} \widehat{k}\widehat{A}\gamma_{5}\gamma_{\mu} + \frac {e}{2kp}\widehat{k}\widehat{A}\gamma_{5}\gamma_{\mu}\frac {e}{2kp}\widehat{A}\widehat{k}\right]u.\eqno(52)$$

Using elementary relations,

$$\bar u u = 2m, u\bar u = 2m, \quad  S_{\mu(0)} = i\bar u \gamma_{5}\gamma_{\mu}u,  \quad v_{\mu} = \bar u
\gamma_{\mu}u,\eqno(53)$$

we get after elementary operations, the following equation for the spin motion 

$$S_{\mu} = i\bar \psi\gamma_{5}\gamma_{\mu}\psi = S_{\mu(0)}\left[1 + \left(\frac {e}{2kp}\right)\frac{1}{2m^{2}}(vk)(vA)
+ \left(\frac
  {e}{2kp}\right)^{2}\frac{1}{16m^{4}}(vk)^{2}(vA)^{2}\right].
\eqno(54)$$

For $A_{\mu} = a_{\mu}\eta(\varphi)$, we get for $\varphi > 0$

$$S_{\mu} = i\bar \psi\gamma_{5}\gamma_{\mu}\psi = i\frac{\bar u}{2p_{0}}
\left[\gamma_{5}\gamma_{\mu} + \frac {e}{2kp}\gamma_{5}\gamma_{\mu}
\widehat{a}\widehat{k} + \frac {e}{2kp} \widehat{k}\widehat{a}\gamma_{5}\gamma_{\mu} + \frac {e}{2kp}\widehat{k}\widehat{a}\gamma_{5}\gamma_{\mu}\frac {e}{2kp}\widehat{a}\widehat{k}\right]u\eqno(55)$$
and

$$S_{\mu} = i\bar \psi\gamma_{5}\gamma_{\mu}\psi = \frac{1}{2p_{0}}S_{\mu(0)} \left[1 + \left(\frac {e}{2kp}\right)\frac{1}{2m^{2}}(vk)(va)
+ \left(\frac
  {e}{2kp}\right)^{2}\frac{1}{16m^{4}}(vk)^{2}(va)^{2}\right].
\eqno(56)$$

There is a surprise that the Wolkow solution for the spin  and vector motion is not identical with the solution following from the BMT equation and Lorentz equation.

\section{Discussion}

We have presented, in this article,
the solution of the BMT equation for spin motion of a charged
particle in the electromagnetic wave of the Dirac $\delta$-function 
pulse field.

The present article is continuation of the author
discussion on electron in laser field
(Pardy, 1998; Pardy, 2001, Pardy 2002) and Lorentz-Dirac equation in delta-fumction pulse (Pardy, 2012),
where the Compton model of laser acceleration
was proposed and quantum theory of motion of electron in laser field was applied.

The $\delta$-form laser pulses are here considered as an idealization of the experimental situation in laser physics. Nevertheless, it was
demonstrated theoretically that at present time the zeptosecond and
subzeptosecond laser pulses of duration $10^{-21} - 10^{-22}$ s can be realized by the petawat lasers. It means that the generation of the ultrashort laser pulses is the keen interest in development of laser physics (Agostini et al., 2004).

Let us remark that while the $\delta$-form pulses are not still used in the theoretical laser physics, such exotic pulses are 
constantly used in the synchrotron physics, where the equation for the betatron radial vibration involves the derivative of the $\delta$ function:

$$r'' + \frac{c^{2}}{R^{2}}r = \frac{1}{E}\sum_{i}\hbar cn
\delta'(t-t_{i}), \eqno(57)$$
where $r$ being the radial deflection from radius $R$ of the local orbit with energy $E$ and the  number of harmonic  $n$.

So the synchrotron theory uses not only $\delta$-form pulses of photons radiated by an electron accelerated on an orbit, but also their derivative (here denoted by symbol $'$).

We have seen that the $\delta$-function form of force is an impact which causes that the body obtains the nonzero velocity or nonzero momentum at $t=0$. The situation in quantum field theory is a such that  $\delta$-function is a source which can generate elementary particles. It is not excluded that the Big Bang started at $t=0$ by  $\delta$-function form of impact. The idea that the existence of  universe started with the zero radius was formulated many years ago by Friedmann and Lama$\hat{\rm i}$tre. 
While the Friedmann
solution follows from the Einstein general relativity, quantum chromodynamics gives  no answer that the Big Bang started by the   
$\delta$-function form  source of quarks and leptons.

New experiments can be realized and new measurements performed by means of the ultrashort laser pulses, giving new results and discoveries. For instance well known transmutation of elements by laser pulse. Specially the photo-desintegration of heavy hydrogen

$$_{1}H^{2} + \gamma  \longrightarrow\; _{1}H^{1} + n\eqno(58)$$
will be replaced in the ELI project by 

$$_{1}H^{2} + \delta{\rm-pulse}  \longrightarrow\;  _{1}H^{1} +
anything
\eqno(59)$$
and in general the following nuclear transmutation will be realized:

$$_{Z}N^{A} + \delta{\rm-pulse}  \longrightarrow\;  _{Z}N^{B} +
anything
\eqno(60)$$
  
So, it is obvious that the interaction of particles with the laser pulses can form, in the near future, the integral part of the laser and particle physics in such laboratories as ESRF, CERN, DESY, SLAC and specially in ELI.

\vspace{10mm}

\noindent
{\bf REFERENCES}

\vspace{7mm}

\noindent
Agostini, P. and DiMauro, L. F. (2004). The physics of attosecond light pulses, Rep. Prog. Phys. {\bf 67}, 813–855. \\[2mm]
Dunne,  M. A. (2006). High-power laser fusion facility for Europe,
Nature Phys. {\bf 2}, 2-5. \\[2mm]
Frenkel, J. I. (1926). Die Elektrodynamik der rotierenden Elektronen, Zs. Physik {\bf 37}, 243. \\[2mm]
Frenkel, J. I. (1958). {\it Collective scientific works}, II., {\it Scientific articles}, AN SSSR, (in Russian).\\[2mm]
Landau, L. D. and  Lifshitz, E. M. (1988). {\it The Classical \\
Theory of Fields}, 4-th ed.~(London, Oxford).\\[2mm]
Marklund, M. and  Shukla, P. K. (2006). Nonlinear collective effects in photon-photon and photon-plasma interactions, Rev. Mod. Phys. {\bf 78}, 591.  \\[2mm]
Martynenko, V. S. (1973). {\it Operator Calculus}, (Kiev, 1973). (in Russian).  \\[2mm]
Mourou, G. A., Tajima, T. and  Bulanov, S. V. (2006). Optics in the relativistic regime Rev. Mod. Phys. {\bf 78}(2), 309. 
\\[2mm]
Meyer, J. W. (1971). Covariant classical motion of electron in a laser beam, Phys. Rev. D {\bf 3}, No. 2, 621 - 622. \\[2mm]
Ohanian, H. C. (1986). What is spin?, Am. J. Phys. {\bf 54}(6), 500. \\[2mm]
Pardy, M. (1973). Classical motion of spin 1/2 particles with zero anomalous magnetic moment,
Acta Phys. Slovaca {\bf 23}, No. 1, 5. \\[2mm]
Pardy, M. (1998). The quantum field theory of laser acceleration,
Phys. Lett. A {\bf 243}, 223-228. \\[2mm]
Pardy, M. (2001). The quantum electrodynamics of laser acceleration,
Radiation Physics and Chemistry {\bf 61}, 391-394.\\[2mm]
Pardy, M. (2012). Lorentz-Dirac equation in the delta-function pulse, arXiv:1208.0488v1 [physics.gen-ph]\\[2mm]
Rafanelli, K, and Schiller, R. (1964). Classical motion of spin-1/2 particles, Phys. Rev.
{\bf 135}, No. 1 B, B279. \\[2mm]
Salamin, Y. I.  Hu, S. X.,  Hatsagortsyan, K. Z. and  Keitel, C. H. (2006). Relativistic high-power laser–matter interactions, Phys. Rep. {\bf 427}(2-3), 41-155. \\[2mm]
Ternov, I. M. (1980). On the contemporary interpretation of the classical theory of the J. I. Frenkel spin, Uspekhi fizicheskih nauk, {}{\bf 132}, 345. (in Russian). \\[2mm]    
Thomas, L. H. (1926). The motion of spinning electron, Nature, {\bf 117}, 514. \\[2mm]
Tomonaga, S.-I. (1997). {\it The story of spin}, 
(The university of Chicago press, Ltd., London).\\[2mm] 
Uhlenbeck, G. E. and Goudsmit, S. A. (1926). Spinning electrons and the structure of spectra, Nature {\bf 117}, 264.\\[2mm] 
Wolkow, D. M. (1935). $\ddot{\rm U}$ber eine Klasse von L$\ddot{\rm o}$sungen der Diracschen Gleichung, Z. Physik, {\bf 94}, 250 - 260.
\\[2mm] 
Yanovsky, V.,  Chvykov, V., Kalinchenko, G., Rousseau, P., Planchon, T., Matsuoka, T., Maksimchuk, A., Nees, J., Cheriaux, G., Mourou, G. and Krushelnick, K. (2008). Ultra-high intensity- 300-TW laser at 0.1 Hz repetition rate,
Optics Express, {\bf 16} Issue 3, 2109-2114. 
\end{document}